\documentclass[5p,twocolumn]{elsarticle}
\usepackage{graphicx,latexsym}
\usepackage{dcolumn}
\usepackage{amssymb,amsmath,bm}
\usepackage{subfigure}
\usepackage{braket}
\usepackage{siunitx}
\usepackage{array}
\usepackage{float}
\usepackage[nopar]{lipsum}
\usepackage{caption}
\usepackage{multirow}
\usepackage{bigstrut}

\biboptions{sort&compress}

\usepackage[super]{nth}

\newcommand{\angstrom}{\text{\normalfont\AA}}
\usepackage{hyperref}
\hypersetup{
    pdfnewwindow=true,       
    colorlinks=true,         
    linkcolor=blue,          
    citecolor=blue,          
    filecolor=magenta,       
    urlcolor=black           
}

\usepackage[normalem]{ulem}

\def\sec#1{Sec.\ \ref{#1}}

\def\fig#1{Fig.\ \ref{#1}}
\def\tab#1{Tab.\ \ref{#1}}

\journal{}

\begin{document}

\begin{frontmatter}



\title{Buckling effects in AlN monolayers: Shifting and enhancing optical characteristics\break from the UV to the near visible light range}

\author[a1,a2]{Nzar Rauf Abdullah}
\ead{nzar.r.abdullah@gmail.com}
\address[a1]{Division of Computational Nanoscience, Physics Department, College of Science, \\
             University of Sulaimani, Sulaimani 46001, Kurdistan Region, Iraq}
\address[a2]{Computer Engineering Department, College of Engineering, Komar University of Science and Technology, \\ Sulaimani 46001, Kurdistan Region, Iraq}

\author[a3]{Botan Jawdat Abdullah}
\address[a3]{Department of Physics , College of Science- Salahaddin University-Erbil, Erbil 44001, Kurdistan Region, Iraq}

\author[a1]{Hunar Omar Rashid}

\author[a5]{Vidar Gudmundsson}
\address[a5]{Science Institute, University of Iceland, Dunhaga 3, IS-107 Reykjavik, Iceland}


\begin{abstract}

The structural, electronic, and optical properties of flat and buckled AlN monolayers are investigated using first-principles approaches. The band gap of a flat AlN monolayer is changed from an indirect one to a direct one, when the planar buckling increases, primarily due to diminishing sp$^2$ overlapping and bond symmetry breaking in the conversion to sp$^3$ bonds.
The sp$^3$ hybridization thus results in a stronger  $\sigma\text{-}\pi$ bond rather than a $\sigma\text{-}\sigma$ covalent bond. The calculations of the phonon band structure indicates that the buckled AlN monolayers are structurally and dynamically stable.
The optical properties, such as the dielectric function, the refractive index, and the optical conductivity of an AlN monolayer are evaluated for both flat systems and those impacted with planar buckling.
The flat AlN monolayer has outstanding optical characteristics in the Deep-UV and absorbs more effectively in the UV spectrum due to its large band gap. The results reveal that optical aspects are enhanced along different directions of light polarization, with a considerable shift in the optical spectrum from Deep-UV into the visible range. Additionally, depending on the polarization direction of the incoming light, increased planar buckling enhances the optical conductivity in both the visible and the Deep-UV domains.
The ability to modify the optical and electronic properties of these essential 2D materials using planar buckling technique opens up new technological possibilities, particularly for optoelectronic devices.

\end{abstract}

\begin{keyword}
Aluminium Nitride monolayer \sep Planar buckling \sep DFT \sep Electronic structure \sep  Optical characteristic
\end{keyword}

\end{frontmatter}

\section{Introduction}
A novel approach for studying the influence of changes in physical properties has emerged with the introduction of 2D materials. A recent trend is to investigate 2D nanomaterials and develop applications for them in practically every field of science \cite{butler2013progress, abdullah2021properties, gupta2015recent, abdullah2022modulation}. Following the discovery of graphene \cite{novoselov2004electric}, various new 2D nanomaterials \cite{mak2016photonics, vogt2012silicene, mannix2018borophene, zhu2015epitaxial} are also being investigated for their unique electrical, optical, mechanical, magnetic, thermal, and catalytic properties. Other monolayers have characteristics that are comparable to, or better, than graphene and can be used in a wide range of technologies and applications \cite{wan2016tuning, khan2020recent, ma2021tunable} because of their extraordinary physical properties and huge specific surface areas.

Other fascinating monolayer materials include group III-Nitride monolayers, like Aluminum Nitride (AlN), which is useful in a range of applications \cite{beshkova2020properties, ben20212d}. AlN monolayers have been the subject of extensive experimental and theoretical investigations. There are a variety of ways to grow or synthesize AlN monolayers, including a vapour phase transport approach, that has been used to create homogenous and smooth AlN monolayer on Si substrates \cite{zhang2007synthesis}, Ultrathin AlN monolayers are epitaxially grown on Ag(111) single crystals using plasma assisted molecular beam epitaxy \cite{tsipas2013evidence}, and the synthesis of an AlN monolayer on Si(111) directed wafer in ammonia molecular beam epitaxy was studied using reflection high energy electron diffraction method \cite{mansurov2015graphene}.

The majority of the 2D monolayer's physical attributes may be modified using various approaches, including doping, strain, electric field, inclusion in heterostructures, and tuning of the planar buckling. For example, doping is one of the strategies used to modify the physical properties of 2D monolayers. DFT has been used to calculate the electronic and optical characteristics of a V doped AlN monolayer in order to improve their properties \cite{javaheri2018electronic}, V atoms are more capable of replacing aluminum atoms with the lowest formation energy, according to structural simulations. When one V atom is replaced, the pure AlN monolayer becomes a p-type semiconductor with metallic characteristics and zero magnetic moment. When two V atoms are substituted, it becomes a half-metallic semiconductor with 100 percent spin polarization. Optical properties such as the dielectric function, the energy loss function, the absorption coefficients, and the refractive index are also estimated for both parallel and perpendicular electric field polarizations, and the findings demonstrate that the optical spectra are anisotropic. The electronic structure of p-type Mg doping in AlN monolayers have been investigated \cite{peng2014tunable}. The findings show that the amount of Mg doped in the AlN monolayer has a significant impact on its structural properties, the formation energies, the transition energy levels, and the bond length between the Mg and the N is found to increase. It has been shown that there are anisotropies in the optical properties of Zr$_2$AlN, with the lowest optical anisotropy in the three principal crystallographic directions of $x$, $y$ and $z$ \cite{YANG2022103962}.

Another factor that affects the characteristics of a 2D monolayer is strain. The influence of strain on the characteristics of AlN monolayers has been investigated. The optical properties of a hexagonal AlN monolayer were examined, and the results indicate that the absorption spectrum is completely redshifted with the applied strain \cite{keccik2015layer}. The effects of strain on the electronic properties of AlN monolayers have been investigated, and it was found that changes in the band structures and the orbital character of the CBM and VBM strongly influence the properties of AlN monolayers \cite{postorino2020strain}. Homogeneous biaxial strain is more efficient than uniaxial one, and shear strain, in modifying the band gap, according to first-principles investigations. The band gap in an AlN monolayer was shifted from an indirect to a direct one due to shear strain \cite{liu2014shear}.

Another major strategy to modify the material characteristics of 2D monolayers is to vary the planar buckling mechanism. A first-principles technique with a high rate of planar buckling has been used to identify the effects of buckling on the electrical and optical characteristics of beryllium oxide monolayers \cite{jalilian2016buckling}. According to previous research into the electrical properties, planar buckling can result in a tunable band gap since the energy band gap decreases when the planar buckling factor is raised. Furthermore, when the planar buckling factor varies, all optical spectra characteristics are redshifted to lower energy, and these shifts can be related to planar buckling weakening bonds and increasing their number.

In our investigation, first-principles density functional theory is used to investigate the electrical and optical properties of a two-dimensional AlN structure. The consequences of buckling on the band structure, which is critical for the electronic structure of the system, are being studied. The band gap of AlN changes dramatically when the distances between the atoms are changed by the buckling. Flat AlN monolayers have an indirect gap with, whereas AlN structures with high buckling rate are direct semiconductors. Moreover, when the planar buckling factor in an optical component is raised, all optical spectra characteristics are shifted to lower energy. Our results significantly contribute information about functionality of novel nanoscale optical device.

The structure of the paper is as follows: \sec{Computational} contains a description of the computational method, and \sec{Results} is a presentation of the calculated electrical and optical properties for AlN monolayers. The last section, \sec{conclusion}, of the paper is the conclusion.

\section{Computational Details}\label{Computational}

Density functional theory, DFT, as applied in the Quantum espresso software package \cite{Giannozzi_2009, giannozzi2017advanced}, is used to execute our computations under the paradigm of first principles. The study considers a two-dimensional AlN monolayer in the  $xy$-plane. The generalized gradient approximation (GGA) is used to compute the band structure, the density of states \cite{ABDULLAH2021106981}, and the optical characteristics of AlN monolayr using the Perdew-Burke-Ernzernhof (PBE) functionals \cite{PhysRevLett.77.3865, doi:10.1063/1.1926272, ABDULLAH2020114556}. In addition, the HSE functional \cite{heyd2004efficient} is employed to obtain accurate estimation of the energy band gaps.

The cutoffs for the energy associated with the charge density, and the kinetic energy of the plane-waves are set to $1.088 \times 10^{4}$~eV, and $1088.5$~eV, respectively, for the sampling of the Brillioun zone for the relaxation of the structure. A vacuum layer is assumed to be $20 \, \angstrom$ in the $z$-direction to avoid possible interaction with the neighboring metallic nitrite layer in this direction. A $18 \times 18 \times 1$ Monkhorst-Pack grid is used to adjust all atomic locations and cell properties until all forces on the atoms are less than $10^{-5}$ eV/$\angstrom$.

For the self consistent field (SCF) and the density of states (DOS) calculations, the same aforementioned parameters are applied, with the exception that the mesh for the momentum space for the DOS calculations is set to  $100 \times 100 \times 1$. By distributing $0.01$ eV tetrahedras, the partial density of states (PDOS) is widened. The QE package with an optical broadening of $0.1$~eV is used to compute the optical characteristics of the AlN monolayers \cite{abdullah2021modulation}.

The phonon calculations are done using the QE software with cutoff energy of $1.156 \times 10^{3}$~eV, and a denser grid of k-mesh points, $22 \times 22 \times 1$. The AlN system is fully relaxed for the phonon calculations with the force less than $10^{-8}$ eV/$\angstrom$ on each atom.

\section{Results and Discussion}\label{Results}

The electrical and optical characteristics of AlN monolayers are studied in this part by examining systems with different values of planar buckling ($\Delta$) with a $2\times2\times1$ supercell monolayer,  where $\Delta$ denotes the buckling strength.

\subsection{Structural properties}

The formation energy, bond length, and electron distribution are investigated here.
We consider the results for the flat AlN monolayer as a reference to compare to for the buckled cases while determining their physical properties. In an AlN monolayer, the Al and the N atoms are in the same plane. When systems with planar buckling are modeled, all the Al atoms are in the same plane, whereas all the N atoms are in a different plane.

 \begin{table}[h]
	\centering
	\begin{center}
		\caption{\label{table_one} The fraction of orbital participation in the bond character and the hybridization bond order in an AlN monolayer.}
		\begin{tabular}{|l|l|l|l|l|l|l|l|}\hline
			$\Delta$ ($\angstrom$)& $s$ ($\%$)  &  $p$ ($\%$) &  Hybrid      \\ \hline
			0.0	    & 33.33 & 66.66 &  sp$^2$       \\
			0.1	    & 33.28 & 66.72 &  sp$^{2.004}$ \\
			0.2	    & 32.48 & 67.52 &  sp$^{2.07}$ \\
			0.3	    & 31.46 & 68.54 &  sp$^{2.17}$ \\
			0.4	    & 30.02 & 69.98 &  sp$^{2.33}$ \\
			0.5	    & 28.16 & 71.84 &  sp$^{2.55}$ \\
			0.6	    & 25.88  & 74.12 &  sp$^{2.86}$  \\
			0.7	    & 23.25  & 76.75  &  sp$^{3.3}$  (not possible) \\ \hline
	\end{tabular}	\end{center}
\end{table}

The allowed values of $\Delta$ are listed in \tab{table_one} for an AlN monolayer. In comparison to the planar AlN monolayer with  $\Delta= 0.0$, we use six different $\Delta$ values ranging from $0.1$ to $0.6$~$\angstrom$, and the crucial value describing no planar buckling.
The maximum allowed value of $\Delta$ is determined by the $s$- and the $p$-orbital hybridization.
The link between the hybridization and the bond angle for $s\text{-}p$ hybrids is determined by this equation,
$\cos(\theta) = s/(s-1) = (p-1)/p$, where $\theta$ is the angle between the respective orbitals and the $s$ and the $p$ parameters are given as decimal fractions \cite{kaufman1993inorganic}. Planar buckling causes the hybridization type to change from sp$^2$ to sp$^3$, according to the findings on hybridization bond order. The orbital orientation and the bond reformation are caused by the hybridization of the bond order. In an AlN monolayer, the case of $\Delta = 0.6$~$\angstrom$ refers to the maximum planar buckling possible. Values greater than this are not possible for the sp$^2$ to sp$^3$ hybridization as is listed in \tab{table_one}.

\begin{table*}[h]
	\centering
	\begin{center}
		\caption{\label{table_two} The formation energy (E$_f$),  of Al-N, and the band gap (E$_g$) using GGA-PBE and HSE06 for different values of planar buckling, $\Delta$.}
		\begin{tabular}{l|l|l|l|l}\hline
			$\Delta$ ($\angstrom$) 	& E$_f$ (eV/atom)  & Al-N ($\angstrom$) & E$_g$ (eV) (GGA-PBE) & E$_g$ (eV) (HSE$06$)   \\ \hline
			0.0	                    & 0.015       & 1.799              &  2.977  &  4.04  \\
			0.1	                    & 0.113       & 1.802              &  2.890  &  3.91  \\
			0.2	                    & 0.453       & 1.810              &  2.674  &  3.73  \\
			0.3	                    & 1.029       & 1.824              &  2.408  &  3.45  \\
			0.4	                    & 1.832       & 1.843              &  2.142  &  3.09  \\
			0.5	                    & 2.883       & 1.868              &  1.902  &  2.87  \\
			0.6	                    & 4.178       & 1.897              &  1.497  &  2.48  \\ \hline
	\end{tabular}	\end{center}
\end{table*}

The formation energy is the initial step in determining the amount of energy necessary to build a structure's atomic configuration.
The formation energy of the AlN monolayers can be defined as
\begin{equation}\label{eq01}
	E_f = E_{\rm Buckling} - E_{\rm Flat},
\end{equation}
where E$_{\rm Buckling}$ is the total energy of the AlN monolayer with planar buckling, and E$_{\rm Flat}$ indicates the total energy of the flat AlN monolayer, without buckling.
In the case of $\Delta=$ 0.0, the formation energy of a flat AlN monolayer is $0.015$~eV.  As planar buckling rises to $0.6$~$\angstrom$, the formation energy increases to $4.178$~eV and the energetic stability decreases. Here, a large formation energy corresponds to a lower structural stability.
To see the formation energy of all values of planar buckling, one may refer to \tab{table_two}. A previous study of BeO layers has shown, that when planar buckling increased, the stability of a BeO monolayer decreased \cite{jalilian2016buckling, ABDULLAH2022106409}.

\begin{figure}[htb]
	\centering
	\includegraphics[width=0.45\textwidth]{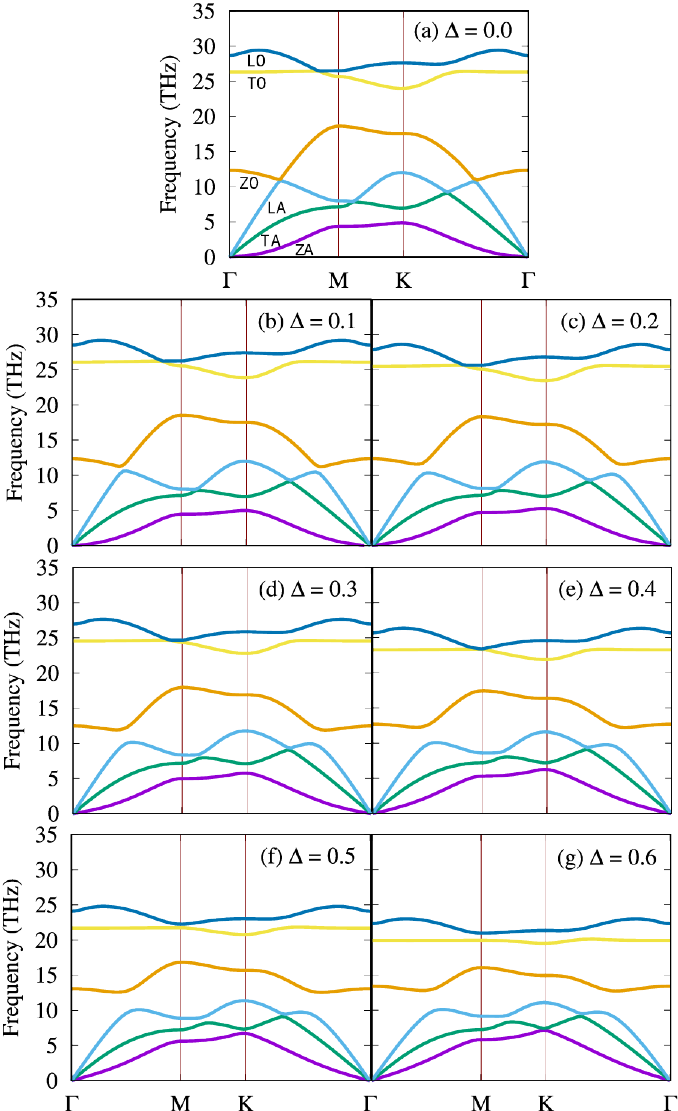}
	\caption{Phonon band structure of an AlN monolayer as a function of the buckling parameter, $\Delta = 0.0$ (a), $0.1$ (b), $0.2$ (c), $0.3$ (d), $0.4$ (e), $0.5$ (f), and $0.6$~$\angstrom$ (g).}
	\label{fig01}
\end{figure}

The dynamical stability of the buckled AlN monolayer can be tested using phonon band structure shown in \fig{fig01}. Unit cells containing two atoms, one Al atom and one N atom, are considered in the phonon calculation.
Consequently, the phonon spectra for the flat and buckled AlN monolayers have 6 phonon branches:
out-of-plane acoustic (ZA), and out-of-plane optical (ZO) modes,
where Z refers to the displacement vector along the $z$ axis; transverse
acoustic (TA), transverse optical (TO), longitudinal acoustic
(LA), and longitudinal optical (LO) modes, indicating vibrations within the 2D plane.
We note that the phonon band structure of the flat AlN monolayer agrees very well with the literature \cite{C4TA03134B}.
The mode frequencies are positive of all buckled AlN monolayers, with no appearance of unphysical imaginary phonon modes, demonstrating that the buckled AlN monolayers are dynamically stable.

Another interesting points to notice are the effects of the buckling on the phonon band structure. Both the acoustic and the optical modes are lowered in energy with increasing $\Delta$. The influence of $\Delta$ is stronger on the optical phonon modes compared to the acoustic phonon modes. This can be explained as follows:
As the buckling parameter is increased, a conversion to an sp$^3$ hybridization is achieved leading to a stronger $\sigma\text{-}\pi$ bond than the $\sigma\text{-}\sigma$ bond. As a result, the tendency of neighboring atoms to oscillate in anti-phase (optical phonon) is more influenced than the in-phase motion of the atoms (acoustic phonon). The lowering of the energy of the phonon modes reflects the reduction of the dynamical stability of a monolayer, but they are still stable as no negative frequencies appear in the calculations of the acoustic modes.

\begin{figure}[htb]
	\centering
	\includegraphics[width=0.22\textwidth]{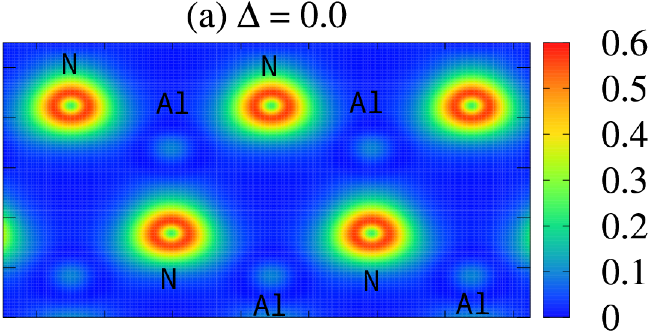}\\
	\includegraphics[width=0.22\textwidth]{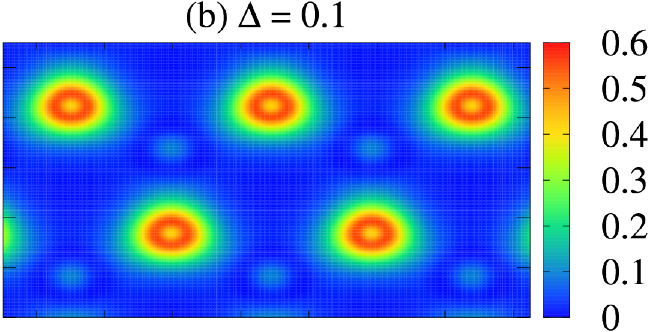}
	\includegraphics[width=0.22\textwidth]{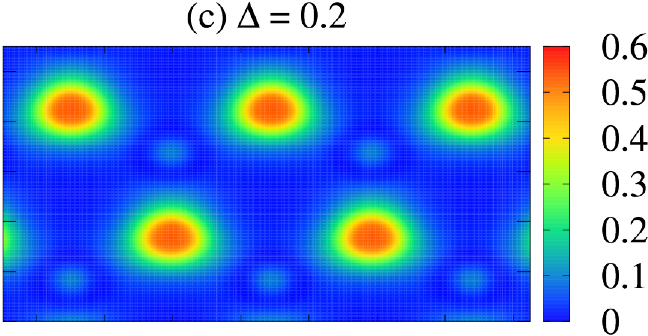}
	\includegraphics[width=0.22\textwidth]{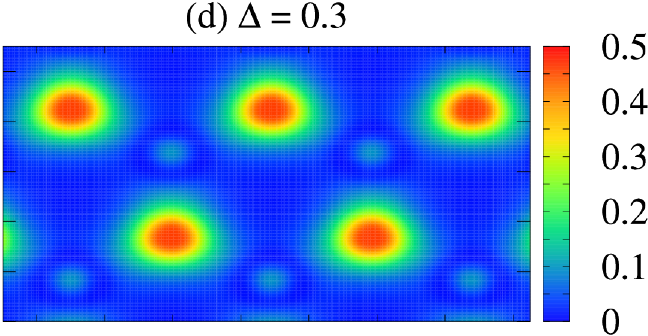}
	\includegraphics[width=0.22\textwidth]{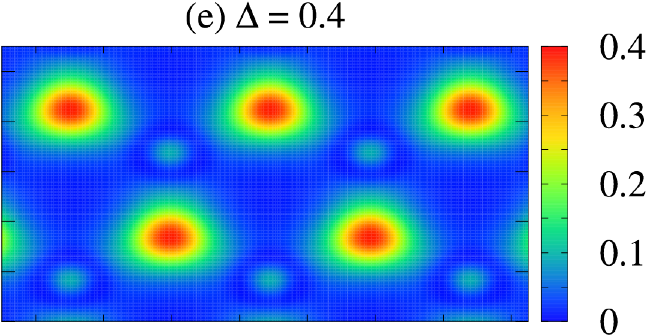}
	\includegraphics[width=0.22\textwidth]{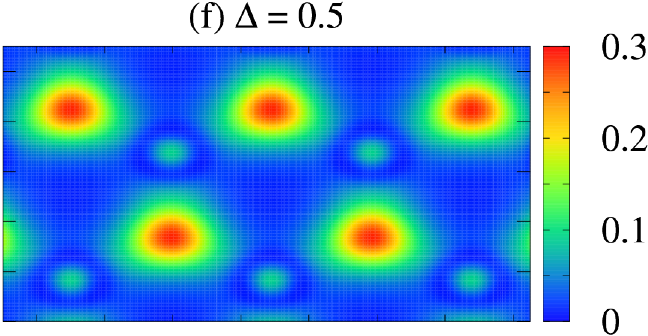}
	\includegraphics[width=0.22\textwidth]{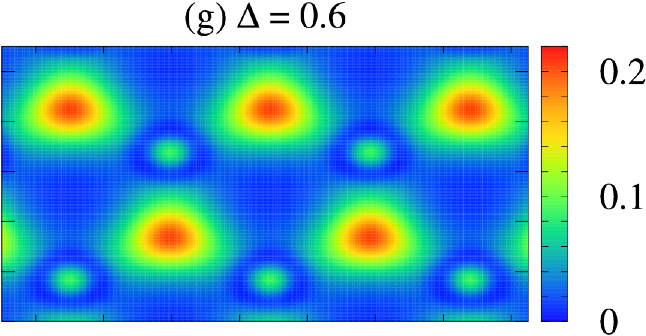}
	\caption{The electron density of an AlN monolayer as a function of the buckling parameter, $\Delta = 0.0$ (a), $0.1$ (b), $0.2$ (c), $0.3$ (d), $0.4$ (e), $0.5$ (f), and $0.6$~$\angstrom$ (g).}
	\label{fig02}
\end{figure}

The electron density is presented in \fig{fig02} displaying the electron density of a flat AlN monolayer as well as the influence of the buckling on the electron density. The electrons in an AlN monolayer are transported from Al to N due to an electronegativity mismatch between the Al and the N atoms. This is not the case for graphene with completely covalent bonds. In an AlN monolayer, the bonding of Al and N takes on an ionic nature.

The N atoms have a larger charge density than the Al atoms due to their higher electronegativity, which is consistent with earlier theoretical studies for AlN monolayers \cite{valedbagi2013electronic, csahin2009monolayer}. A strong $\sigma\text{-}\sigma$ bond forms between the Al and N atoms as a result of the sp$^2$ hybridization. The electron density distribution around both the Al and N atoms is changed when $\Delta$ increases due to the elongation of their bonds as presented in \tab{table_two}.
The bond length elongation leads to less charge transferred from the Al to the N atoms.
We thus see less electron density around the N atoms and a higher electron density around the Al atoms at higher values of planar buckling, $\Delta = 0.6$~$\angstrom$.

\subsection{Band structure and density of states}

\begin{figure}[htb]
	\centering
	\includegraphics[width=0.47\textwidth]{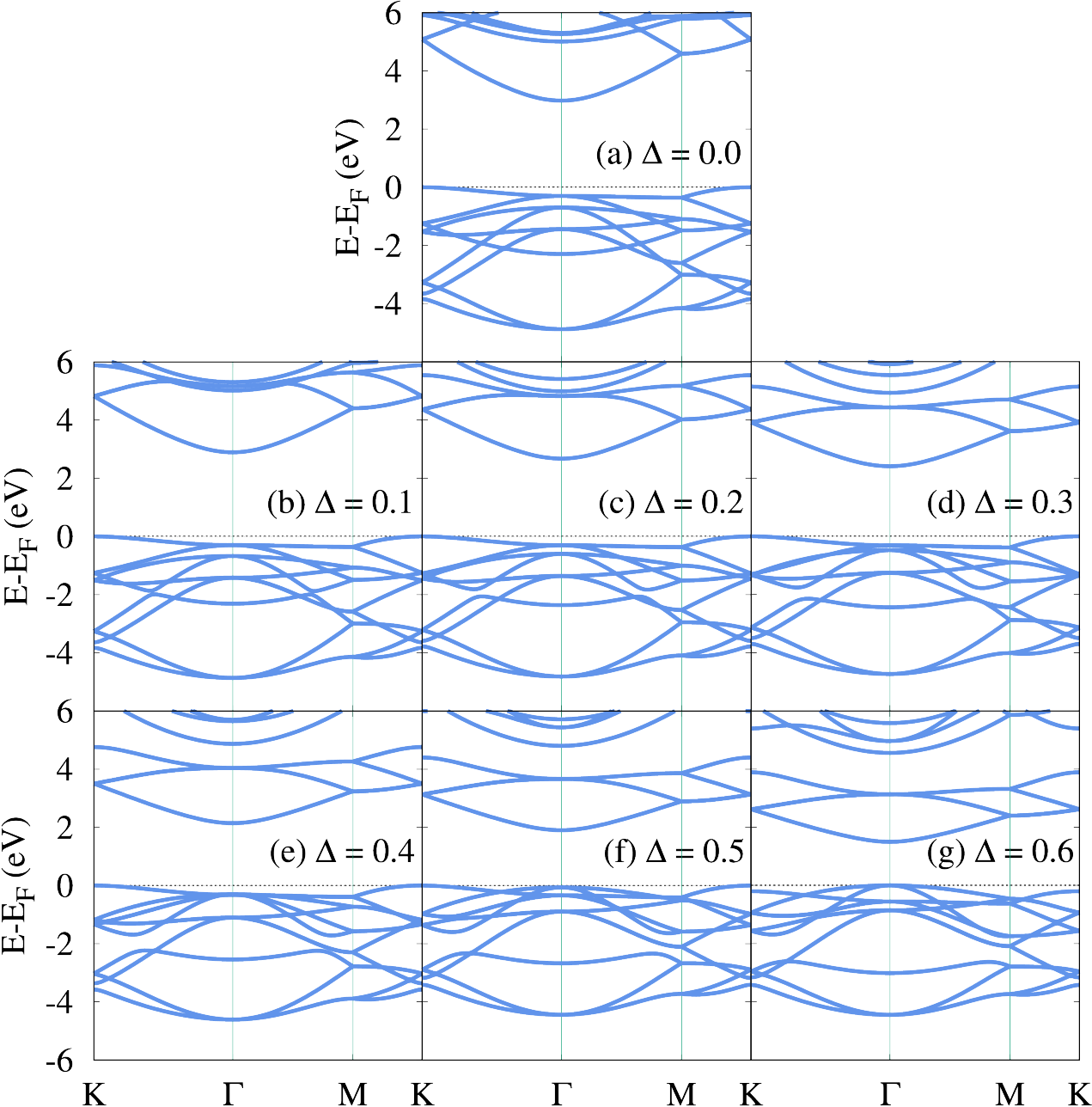}
	\caption{Band structure for optimized AlN monolayer with buckling parameter,  $\Delta = 0.0$ (a), $0.1$ (b), $0.2$ (c), $0.3$ (d), $0.4$ (e), $0.5$ (f), and $0.6$~$\angstrom$ (g).
		The energies are with respect to the Fermi level, and the Fermi energy is set to zero.}
	\label{fig03}
\end{figure}

The bond length modification due to the planar buckling break the symmetry of an AlN monolayer, and thus the electronic band structure. The band structure of a flat AlN monolayer without buckling is seen in \fig{fig03}, as well as the effect of the buckling on the band structure. The band gap of a flat AlN monolayer is determined to be  $2.977$ eV within the GGA along the $\Gamma$-K pathway, which coincides well with previous DFT investigations \cite{luo2020novel, valedbagi2013electronic}. Because the band gap within the GGA is underestimated, the HSE is also employed to obtain more accurate band gap data. In the case of a flat AlN monolayer, the band gap is equal to $4.04$ eV using the HSE$06$ method (see \tab{table_two}). As the planar buckling is increased, the bond lengths of the AlN monolayer expand. As a result, the band gap of the AlN monolayer is significantly lowered for greater values of $\Delta$.
The longer bond lengths shows that the electrons are less tightly bound to the
atoms. In this case, less energy is required to remove an electron from the atoms
leading to a decreased band gap \cite{gacevic:hal-00632224, ABDULLAH2022106835}. We thus see the band gap of an AlN monolayer with planar buckling of $0.6$~$\angstrom$ is $1.487$~eV, which is almost half of the band gap of the flat AlN monolayer (see \tab{table_two}). As is seen from \tab{table_two} the increased planar buckling causes the Al-N bond length to approach the Al-N bond length of bulk AlN, which is $1.90 \, \angstrom$ \cite{PhysRevMaterials.2.064607}.

\begin{figure}[htb]
	\centering
	\includegraphics[width=0.4\textwidth]{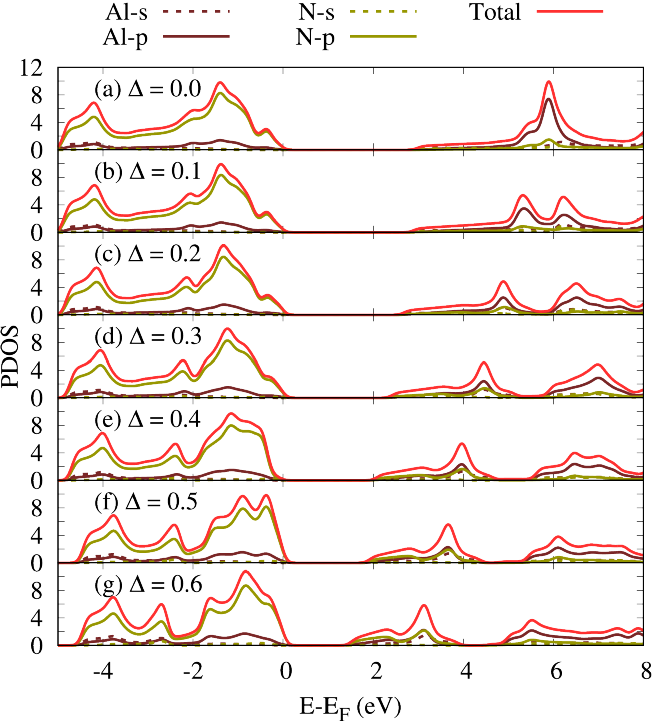}
	\caption{Partial density of states, PDOS of the AlN monolayer with buckling parameter, $\Delta = 0.0$ (a), $0.1$ (b), $0.2$ (c), $0.3$ (d), $0.4$ (e), $0.5$ (f), $0.6$~$\angstrom$ (g).
		The energies are with respect to the Fermi level, and the Fermi energy is set to zero.}
	\label{fig04}
\end{figure}

Additionally, the flat AlN monolayer has an indirect band gap, as shown in the \fig{fig03}. When planar buckling increases, however, the AlN monolayer yields a direct band gap occurring at $\Delta=0.6$~$\angstrom$, by moving the conduction band minima closer to the valence band maxima. The main reason for this is that when $\Delta$ increases, the $\sigma\text{-}\sigma$ covalent bond weakens in relation to the creation of a sp$^3$ hybridization leading to a stronger $\sigma\text{-}\pi$ bond in the AlN monolayer. Similar trend of variation was found in a GeC monolayer in a previous study \cite{ABDULLAH2023107163}.

The band gap reduction can be further explored by presenting the partial density of states (PDOS) as given in \fig{fig04}, where the PDOS of both flat and planar AlN monolayers is shown.
Our findings show that the $p$-orbitals of the N atoms, the N-$p$ states, control the valence band in a flat AlN monolayer, with a small contributions from the $p$-orbitals of the Al atoms, the Al-$p$ states, and and very little of the $s$-orbitals of the Al atoms, the Al-$s$ states.
In contrast, the conduction band is mostly made up of the Al-$p$ states, although there is also a small contribution from the N-$p$ and the Al-$s$ states.
However, when $\Delta$ increases, the density of states becomes more sensitive, and the valance band broadens, but no shift is seen. Additionally, the $p$-orbitals of the Al atoms are gradually shifted downwards into the conduction band near the Fermi energy until they align with the $s$-orbitals, and the $\sigma\text{-}\sigma$ covalent bonds break. Consequently, the band gap reduction is caused by the planar buckling described by $\Delta$.

\subsection{Optical properties}

\begin{figure*}[htb]
	\centering
	\includegraphics[width=0.8\textwidth]{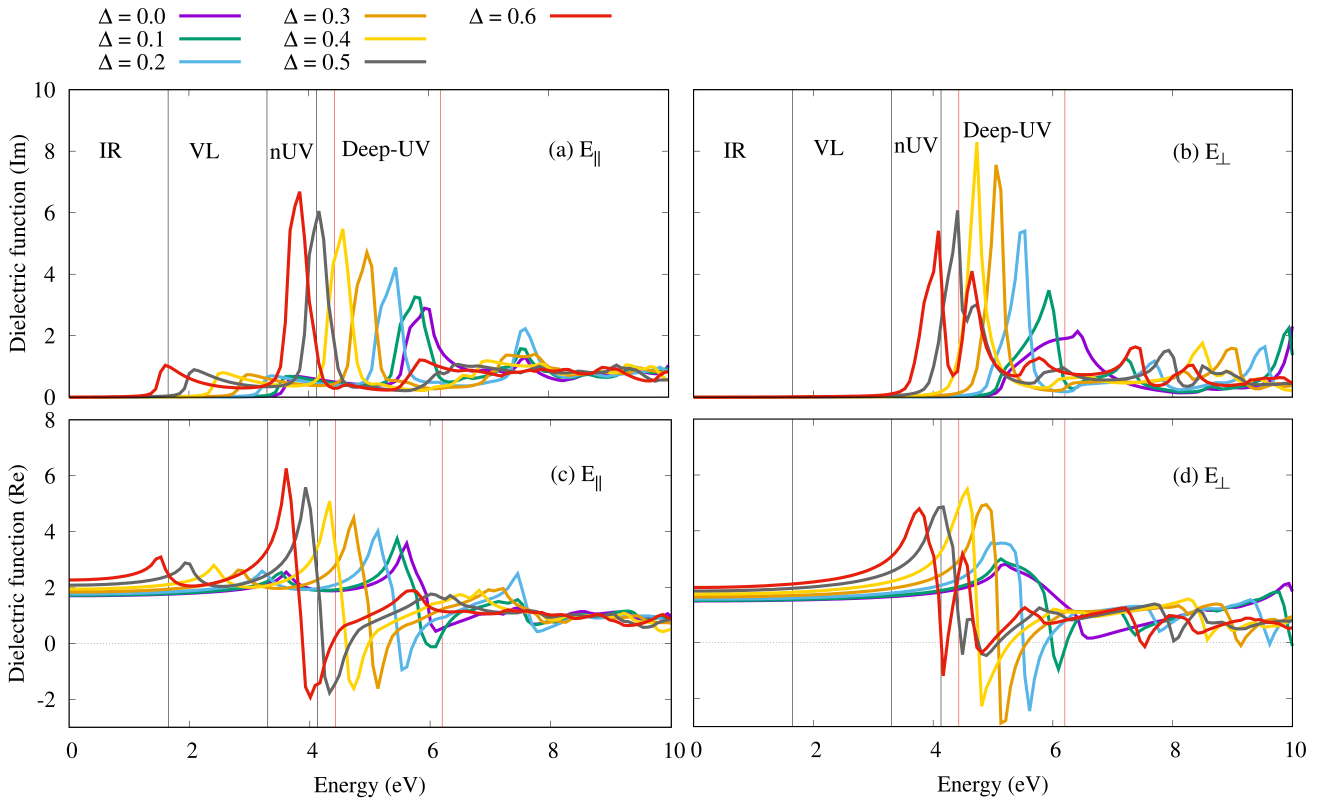}
	\caption{Imaginary, Im($\varepsilon$), (a,b) and real, Re($\varepsilon$), (c,d) parts of the dielectric function for the AlN monolayer with buckling parameter, $\Delta = 0.0$ (purple), $0.1$ (green), $0.2$ (light blue), $0.3$ (orange), $0.4$ (yellow), $0.5$ (gray), and  $0.6$~$\angstrom$ (red) in the case of E$_{\parallel}$ (left panel), and E$_{\perp}$ (right panel). The vertical black lines show different regions of the electromagnetic spectrum.}
	\label{fig05}
\end{figure*}

We investigate here how planar buckling affects the optical characteristics of an AlN monolayer. To compute the excitation energy, the static dielectric function, the refractive index, and the optical conductivity, the random phase approximation (RPA) is applied \cite{ren2012random}. The RPA in the QE package uses a dense $100 \times 100 \times 1$ mesh grid in the Brillouin zone \cite{PhysRev.115.786} to provide high accuracy. The \fig{fig05} represents the imaginary, Im($\varepsilon$), and the real, Re($\varepsilon$), components of the dielectric function for parallel (E$_{\parallel}$) and perpendicular (E$_{\perp}$) polarization of the incoming electric field.

In the case of E$_{\parallel}$, a significant peak at $5.93$~eV of Im($\varepsilon$) can be seen in the Deep-UV region when a parallel electric field is applied to a flat AlN monoalayer (see \fig{fig05}a), which is similar to previous findings for a flat AlN monolayer \cite{valedbagi2013electronic, fakhrabad2015optical}. The peak appearing in the Deep-UV region is related to the indirect band gap of flat AlN monolayer.
In addition, a small peak at $3.76$~eV in Im($\varepsilon$) is found in the near-UV (nUV) corresponding to the optical band gap of a flat AlN monolayer indicating the electron excitation from the valance to the conduction band.
Both peaks are transferred to lower energy as $\Delta$ increases. The main peak is shifted down to the nUV from the Deep UV, and the peak corresponding to the optical band gap is shifted into the visible (VL) and Infrared (IR) regions. The shifted peaks indicate the formation of more states closer to the Fermi energy.

Compared to a flat AlN monolayer with higher $\Delta$, the peaks are shifted to lower energy starting in the Deep-UV and extending to the visible area, and even IR for $\Delta=0.6$. This is referred to the change of the indirect band gap of a flat AlN to a direct one, when the planar buckling is increased. This is useful for optoelectronic devices that function in the visible light spectrum. Due to the change in the band structure and the formation of a direct band gap, the intensity of the peak also increased. A notable increase in Im($\varepsilon$) is seen in the identified electromagnetic spectrum regions for the case of (E$_{\perp}$). Apart from a high peak indicating that the optical band gap shifts from the Deep-UV to the nUV regimes as $\Delta$ increases for (E$_{\perp}$), there are no significant small peaks at the lower energy.

The real part of the dielectric function connects to several interesting physical phenomena such as the static dielectric constant, and the plasmon energy of the AlN monolayer.
The real component of the dielectric function at zero energy, also known as the static dielectric constant Re$(\varepsilon(0))$, is another significant quantity to consider. The value of Re$(\varepsilon(0))$ is increased with $\Delta$, which is expected as the value of Re$(\varepsilon(\omega))$ is inversely proportional to the band gap, Re$(\varepsilon(\omega)) \approx 1/E_{g}$ \cite{PhysRev.128.2093}.
The band gap is decreased with $\Delta$, and the value of Re$(\varepsilon(0))$ is thus increased.
The increase in Re$(\varepsilon)$ for E$_{\parallel}$ is greater than for E$_{\perp}$, indicating that Re$(\varepsilon)$ has an asymmetric behavior.

Plasmons, which are linked to the real component of the dielectric function, are another important phenomena. Collective oscillations of valence or conduction electrons in a material are known as plasmons. This is typically described by a material's complex dielectric function, in which the real component describes electromagnetic wave transmission through the medium and the imaginary component describes single particle excitation, which are characterized by interband transitions \cite{raether2006excitation, egerton2011introduction}. It is known that at the plasmon frequency, the imaginary part of dielectric function has to be maximum and the real part of dielectric function is zero.
It means that the real part of dielectric function changes its sign from positive to negative at the plasmon energy.
The plasmon energy for the flat AlN monolayer is thus seen in the Deep-UV region. However,  when $\Delta$ increases, the plasmon energy is shifted to a lower energy scale from the Deep-UV to the nUV as is shown in \fig{fig05}.

\begin{figure}[htb]
	\centering
	\includegraphics[width=0.4\textwidth]{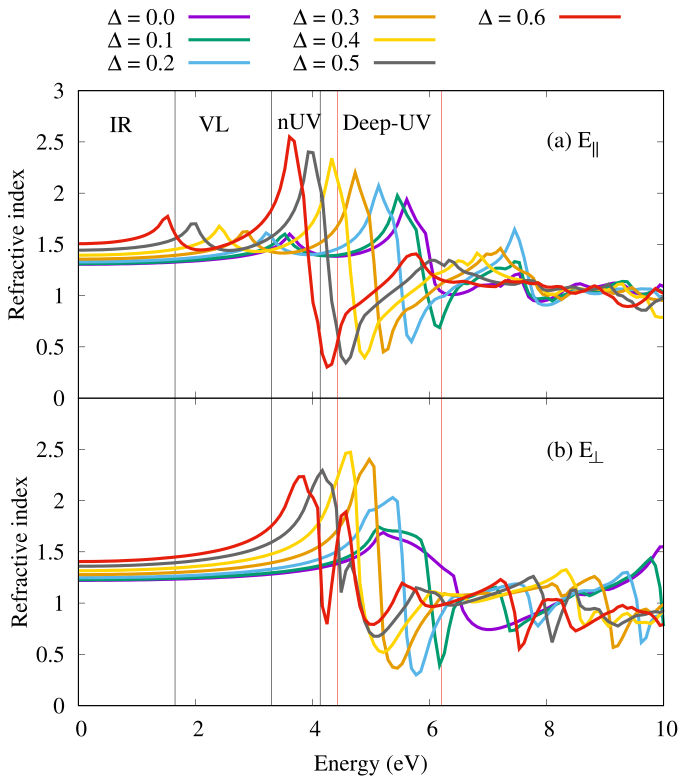}
	\caption{Refractive index, $n(\omega)$ for the AlN monolayer with buckling parameter, $\Delta = 0.0$ (purple), $0.1$ (green), $0.2$ (light blue), $0.3$ (orange), $0.4$ (yellow), $0.5$ (gray), and  $0.6$~$\angstrom$ (red) in the case of E$_{\parallel}$ (a), and E$_{\perp}$ (b). The vertical black lines show different regions of the electromagnetic spectrum.}
	\label{fig06}
\end{figure}

The refractive index $n(\varepsilon)$ of an AlN monolayer versus energy for both polarizations in a flat system and layers with variable planar buckling is shown in the \fig{fig06}. It is obvious that the characteristic of the refractive index follows the real part of the dielectric function.
In parallel polarization, the static refractive index n(0) of a flat AlN monolayer is 1.3, whereas in perpendicular polarization, it is 1.2. However, for both E$_{\parallel}$ and E$_{\perp}$, the $n(0)$ increases as $\Delta$ is increased. The effect of planar buckling tunes the spectrum of $n(\varepsilon)$, and the change follows the trend of the change in the real dielectric function, with the strong peak shifting from the Deep-UV to the nUV area and the weak peak shifting from the Deep-UV to the VL region.

\begin{figure}[htb]
	\centering
	\includegraphics[width=0.4\textwidth]{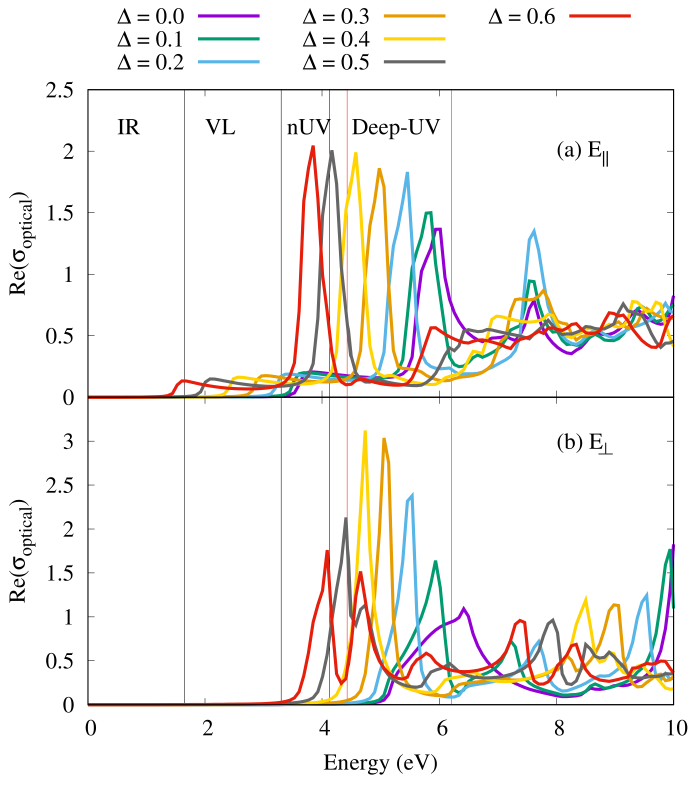}
	\caption{Real part of optical conductivity, $\mathrm{Re}(\sigma_{\rm optical})$, for the AlN monolayer with buckling parameter, $\Delta = 0.0$ (purple), $0.1$ (green), $0.2$ (light blue), $0.3$ (orange), $0.4$ (yellow), $0.5$ (gray), and $0.6$~$\angstrom$ (red) in the case of E$_{\parallel}$ (a), and E$_{\perp}$ (b). The vertical black lines show different regions of electromagnetic wave.}
	\label{fig07}
\end{figure}

Finally, \fig{fig07} shows the real component of the optical conductivity spectra, $\sigma_{\rm optical}$, estimated from the electronic band structure for E$_{\parallel}$ (a) and E$_{\perp}$ (b), respectively, for buckled and planar AlN monolayers. For AlN monolayers with flat structures below $3.52$ eV (E$_{\parallel}$) and $5.0$ eV (E$_{\perp}$), $\sigma_{\rm optical}$ is zero.
In the case of E$_{\parallel}$, the far side of the nUV spectrum has a small peak, whereas the Deep-UV area shows a strong peak. Then another peak appears above $7.0$~eV, this one is related to a higher energy levels. The intensity of the small peak is almost unchanged with $\Delta$, but its location dislocates to lower energy indicating the optical band gap. The intensity of the peak appearing in the Deep-UV for the flat AlN monolayer is prominently enhanced with $\Delta$ and shifts to the nUV region.
Furthermore, the polarization direction of the incoming light has a significant influence on the optical conductivity when the planar buckling is increased. In the case of E$_{\perp}$, the main peak in the Deep-UV region begins to drop in energy as the planar buckling is increased, resulting in a sharper, but weaker intensity.

In general, the strong peak inside the Deep-UV area starts to drop and move to lower energy for both directions of the light polarization, remaining in the UV zone for the high buckling before shifting to the near UV region. When the photon energy grows and exceeds the energy gap, a new conduction mechanism can arise. The real part of the conductivity has a frequency dependence due to an interband transition. As a phonon is involved in the indirect transitions, an increased planar buckling changes the band structure facilitating direct transitions with the crystal momentum retained without a large change in the wave vector.

\section{Conclusions}\label{conclusion}

We have used density functional theory to investigate the effects of a planar buckling on the electrical structure and the optical characteristics of an AlN monolayer. The alterations of the band structure and the band gap shifting from an indirect one to a direct one as the planar buckling is increased are studied. The planar buckling affects the contributions to the density of states, and can be used to tune it. The N atoms have a greater charge density than the Al atoms in a flat AlN monolayer due to their higher electronegativity, however the electron density distribution around the N and the Al atoms shifts. It decreases around the N atoms, while being enhanced around the Al atoms as a result of increasing bond lengths.
Parallel and perpendicular polarization of incoming electric fields with respect to the plane of the AlN monolayer are employed to explore the optical characteristics of the AlN monolayer. With increasing buckling strength, the optical spectra are enhanced for both of these polarizations. The two main peaks of the real and the imaginary parts of the dielectric function are shifted from the Deep-UV to the near visible area owing to the appearance of $\sigma\text{-}\pi$ bonds in the sp$^3$ hybridization. Small peaks are related to the electronic band gap and the system remains a semiconductor as the band gap narrows, but strong peaks are associated with interband transitions owing to the frequency dependence of the real component of the conductivity.  As a result, the optical characteristics can be tuned by changing the planar buckling in the Deep-UV region.
A flat AlN monolayer has excellent optical characteristics in the Deep-UV and absorbs more effectively in the UV spectrum due to its large band gap, while the region of the absorption is modified by increased planar buckling from the UV to the VL region, reflecting a stronger  $\sigma\text{-}\pi$ bond. The findings are of importance for optoelectronic technological development and devices.

\section{Acknowledgment}
This study was funded in part by the University of Sulaimani and the Komar University of Science and Technology's Research Center. The simulations were carried out utilizing resources provided by the University of Sulaimani's department of Computational Nanoscience.



\end{document}